\documentclass[12pt,eqno,epsf]{article}
\usepackage{amsmath,amssymb,graphicx,slashbox}
\numberwithin{equation}{section}

\def\mydate{October 15, 2016}
\def\ignore#1{{}}

\newcounter{sxn}

\newcounter{axn}

\date{}

\newdimen\mybaselineskip
\renewcommand{\thefootnote}{\arabic{footnote}}

\newcommand{\beeq}{\begin{equation}}
\newcommand{\eneq}{\end{equation}}
\newcommand{\beqn}{\begin{eqnarray}}
\newcommand{\eeqn}{\end{eqnarray}}

\newcommand{\alp}{\alpha}

\newcommand{\gm}{\gamma}
\newcommand{\Gm}{\Gamma}
\newcommand{\dlt}{\delta}

\newcommand{\ep}{\epsilon}

\newcommand{\tht}{\theta}

\newcommand{\lmd}{\lambda}
\newcommand{\Lmd}{\Lambda}
\newcommand{\sgm}{\sigma}

\newcommand{\vph}{\varphi}

\newcommand{\be}{\begin{equation}}
\newcommand{\ee}{\end{equation}}
\newcommand{\bea}{\begin{eqnarray}}
\newcommand{\eea}{\end{eqnarray}}
\newcommand{\eql}{\!\!\!&=\!\!\!&}

\newcommand{\defa}{\!\!\!&\equiv\!\!\!&}

\newcommand{\toa}{\!\!\!&\to\!\!\!&}

\newcommand{\simgt}{\stackrel{>}{{}_\sim}}
\newcommand{\simlt}{\stackrel{<}{{}_\sim}}

\newcommand{\tl}[1]{\tilde{#1}}
\newcommand{\bdm}[1]{{\mbox{\boldmath $#1$}}}

\newcommand{\der}{\partial}
\newcommand{\dr}{\!\!d}

\newcommand{\ie}{{i.e.}}
\newcommand{\id}{\mbox{\boldmath $1$}}

\newcommand{\cm}{c}
\newcommand{\hcm}{\hat{c}}

\newcommand{\vev}[1]{\langle #1 \rangle}

\newcommand{\brkt}[1]{\left( #1 \right)}
\newcommand{\brc}[1]{\left\{ #1 \right\}}
\newcommand{\sbk}[1]{\left[ #1 \right]}
\newcommand{\abs}[1]{\left| #1 \right|}

\renewcommand{\Re}{{\rm Re}\,}
\renewcommand{\Im}{{\rm Im}\,}

\newcommand{\cC}{{\cal C}}

\newcommand{\cL}{{\cal L}}

\newcommand{\cO}{{\cal O}}

\begin{document}
\thispagestyle{empty}

\baselineskip=12pt

{\small \noindent \mydate    
\hfill }

{\small \noindent \hfill KEK-TH-1917}

\baselineskip=35pt plus 1pt minus 1pt

\vskip 1.5cm

\begin{center}
{\Large\bf Spectrum in the presence of brane-localized mass}\\
{\Large\bf on torus extra dimensions}

\vspace{1.5cm}
\baselineskip=20pt plus 1pt minus 1pt

\normalsize

{\bf Yutaka\ Sakamura}${}^{1,2}\!${\def\thefootnote{\fnsymbol{footnote}}
\footnote[2]{\tt e-mail address: sakamura@post.kek.jp}}

\vspace{.3cm}
${}^1${\small \it Department of Particles and Nuclear Physics, \\
SOKENDAI (The Graduate University for Advanced Studies), \\ 
Tsukuba, Ibaraki 305-0801, Japan} \\ \vspace{3mm}
${}^2${\small \it KEK Theory Center, Institute of Particle and Nuclear Studies, 
KEK, \\ Tsukuba, Ibaraki 305-0801, Japan} 
\end{center}

\vskip 1.0cm
\baselineskip=20pt plus 1pt minus 1pt

\begin{abstract}
The lightest mass eigenvalue of a six-dimensional theory compactified on 
a torus is numerically evaluated in the presence of the brane-localized mass term. 
The dependence on the cutoff scale~$\Lmd$ is non-negligible 
even when $\Lmd$ is two orders of magnitude above the compactification scale, 
which indicates that the mass eigenvalue is sensitive to the size of the brane,  
in contrast to five-dimensional theories. 
We obtain an approximate expression of the lightest mass in the thin brane limit, 
which well fits the numerical calculations,  
and clarifies its dependence on the torus moduli parameter~$\tau$. 
We find that the lightest mass is typically much lighter than 
the compactification scale by an order of magnitude 
even in the limit of a large brane mass. 
\end{abstract}


\newpage

\section{Introduction}
Many extra-dimensional models have four-dimensional (4D) brane-like defects 
on the compact space, such as orbifold fixed points 
or solitonic objects~\cite{ArkaniHamed:1998rs}-\cite{Randall:1999vf}. 
We can freely introduce 4D terms localized 
at the branes\footnote{
Here we do not consider branes spread over other dimensions. 
The word~``brane'' is understood as the ``3-brane'' in this paper. 
}~\cite{Goldberger:2001tn,Davoudiasl:2002ua,delAguila:2006atw}. 
Such brane-localized terms are induced by quantum effect 
even if they are absent at tree level~\cite{Mirabelli:1997aj,Georgi:2000ks}. 
They change the Kaluza-Klein (KK) spectrum 
and deform the profiles of the mode 
functions~\cite{Cacciapaglia:2005da,Dudas:2005vn,Maru:2010ap}. 
In particular, the brane-localized mass terms are often introduced 
in order to remove unwanted modes from the 4D effective 
theory~\cite{Agashe:2006at,Hosotani:2007qw,Hosotani:2008tx}. 
In five-dimensional (5D) theories, the effects of such brane masses 
can be translated into the change of the boundary conditions 
for the bulk fields. 
This is because the branes in 5D can be regarded as the boundaries of the extra dimension. 
In this case, 
large brane masses can make zero-modes of the bulk fields heavy enough 
up to half of the compactification scale. 

In contrast, the branes are no longer the boundaries of the extra compact space 
in higher-dimensional theories. 
Since effects of the brane terms spread over higher-dimensional space 
and are diluted, they are expected to be smaller than those in the 5D case. 
Therefore, it is important to check whether the brane mass can make 
unwanted modes heavy enough or not. 
In this paper, we evaluate the lightest mass eigenvalue of 
a six-dimensional (6D) theory in the presence of the brane-localized mass term. 
The authors of Ref.~\cite{Dudas:2005vn} discussed a closely related issue 
in the case of the $T^2/Z_2$ compactification 
whose torus moduli parameter is $\tau=i$, and obtained 
the result that the inverse of the lightest 
mass eigenvalue has a logarithmic dependence on the cutoff scale. 
Here we generalize their setup and consider a generic torus whose moduli parameter 
is arbitrary. 
Then we can explicitly see the relation to the well-known results in the 5D theories 
by squashing or stretching the torus. 
Besides, we are interested in a different parameter region 
from that discussed in Ref.~\cite{Dudas:2005vn}. 
We mainly focus on the limit of a large brane mass, 
in which the dependence of the mass eigenvalues on the brane mass is negligible, 
and evaluate the ratio of the lightest mass to the compactification scale 
by numerical calculations.

The paper is organized as follows. 
After explaining the setup in the next section, 
we will see the dependences of the lightest mass eigenvalue 
on the cutoff scale of the theory and on the brane mass in Sec.~\ref{MassMatrix}. 
In Sec.~\ref{ap_expr}, we find an approximate expression of the lightest mass 
as a function of the torus moduli parameter~$\tau$, 
and estimate its ratio to the compactification scale. 
Sec.~\ref{summary} is devoted to the summary. 
We provide a brief review of the case of a 5D theory in Appendix~\ref{5Dcase}, 
and discuss theories with fermion or vector field in Appendix~\ref{other_cases}.

\section{Setup} \label{setup}
We consider a 6D theory of a complex scalar field~$\phi$ as a simple example.~\footnote{
Cases of fermion and vector fields are briefly discussed in Appendix~\ref{other_cases}. 
}
The Lagrangian is given by
\be
 \cL = -\der^M\phi^*\der_M\phi-\cm^2\abs{\phi}^2\dlt(x^4)\dlt(x^5)+\cdots, 
 \label{cL}
\ee
where $M=0,1,2,\cdots,5$, and the ellipsis denotes interaction terms, 
which are irrelevant to the following discussion. 
The brane mass parameter~$\cm$ is a real dimensionless constant. 
The extra dimensions are compactified on a torus~$T^2$.\footnote{
The spectrum in the case of $T^2/Z_N$ compactification ($N=2,3,4,6$) 
can easily be obtained by thinning out the spectrum on $T^2$. 
} 
The background metric is assumed to be flat, for simplicity. 
For the coordinates of the extra dimensions, it is convenient to use a complex 
(dimensionless) coordinate~$z\equiv\frac{1}{2\pi R}(x^4+ix^5)$, 
where $R>0$ is one of the radii of $T^2$. 
The torus is defined by identifying points in the extra dimensions as
\be
 z \sim z+n_1+n_2\tau, \;\;\; (n_1, n_2 \in \mathbb{Z})
\ee
where $\tau$ is a complex constant that satisfies $\Im\tau>0$. 

The Lagrangian~(\ref{cL}) is then rewritten as
\be
 \cL = -\der^\mu\phi^*\der_\mu\phi-\frac{1}{2(\pi R)^2}\brc{
 \abs{\der_z\phi}^2+\abs{\der_{\bar{z}}\phi}^2+\cm^2\abs{\phi}^2\dlt^{(2)}(z)}+\cdots,  
 \label{cL2}
\ee
where $\mu=0,1,2,3$, and we have used that
\be 
 \dlt(x^4)\dlt(x^5) = \frac{1}{2(\pi R)^2}\dlt^{(2)}(z). 
\ee

We can expand $\phi$ as  
\be
 \phi(x^\mu,z) = \sum_{n,l=-\infty}^\infty f_{n,l}(z)\phi_{n,l}(x^\mu), 
 \label{KKexpand}
\ee
where 
\be
 f_{n,l}(z) = \frac{1}{2\pi R\sqrt{\Im\tau}}\exp\brc{
 \frac{2\pi i}{\Im\tau}\Im\brc{(n+l\bar{\tau})z}} 
\ee
are normalized as
\bea 
 &&\int_{T^2}dx^4dx^5\;\abs{f_{n,l}\brkt{\frac{x^4+ix^5}{2\pi R}}}^2 
 = 2(\pi R)^2\int\dr^2z\;\abs{f_{n,l}(z)}^2 \nonumber\\
 \eql (2\pi R)^2\Im\tau\int_0^1 dw_1\int_0^1 dw_2\;
 \abs{f_{n,l}(w_1+\tau w_2)}^2 = 1, 
\eea
and satisfy 
\be
 \der_z\der_{\bar{z}}f_{n,l} = -\tl{\lmd}_{n,l}^2f_{n,l},  \;\;\;\;\;
 \tl{\lmd}_{n,l} = \frac{\pi\abs{n+l\tau}}{\Im\tau}.  \label{wocm}
\ee
This corresponds to the KK expansion 
in the absence of the brane-localized mass term. 
The KK masses are given by $\tl{m}_{n,l}\equiv \tl{\lmd}_{n,l}/(\pi R)$. 

Since the 6D theory is non-renormalizable, it should be regarded as 
an effective theory valid only below the cutoff scale~$\Lmd$. 
Here we relabel the KK modes by using the KK label~$a=0,1,2,\cdots$ 
defined in such a way that 
\be
 0 = \tl{m}_0 < \tl{m}_1 \leq \tl{m}_2 \leq \cdots \leq \tl{m}_{N_\Lmd}
 < \Lmd \leq \tl{m}_{N_\Lmd+1} \leq \cdots. 
\ee
The correspondence of the labels~$(n,l)$ and $a$ depends on the value of $\tau$,  
as shown in Tables~\ref{relabelKK:1} and \ref{relabelKK:2}. 
\begin{table}[t]
\begin{center}
\begin{tabular}{|c||c|c|c|c|c|c|c|c|}
 \hline \rule[-2mm]{0mm}{7mm}  
 $a$ & 0 & 1 & 2 & 3 & 4 & 5 & 6 & 7 \\ \hline
 $(n,l)$ & (0,0) & $(1,0)$ & $(0,1)$ & $(0,-1)$ & $(-1,0)$ & $(1,1)$ & $(1,-1)$ & 
 $(-1,1)$ \\ \hline 
 $\tl{\lmd}_a$ & 0 & 3.14 & 3.14 & 3.14 & 3.14 & 4.44 & 4.44 & 4.44 
 \\\hline\hline 
 $a$ & 8 & 9 & 10 & 11 & 12 & 13 & 14 & $\cdots$ \\ \hline
 $(n,l)$ & $(-1,-1)$ & (2,0) & (0,2) & $(0,-2)$ & $(-2,0)$ & (2,1) & $(2,-1)$ & 
 $\cdots$ \\ \hline
 $\tl{\lmd}_a$ & 4.44 & 6.28 & 6.28 & 6.28 & 6.28 & 7.02 & 7.02 & $\cdots$ \\ \hline
\end{tabular}
\end{center}
\caption{Relabeling the KK modes in the case of $\tau=i$}
\label{relabelKK:1}
\begin{center}
\begin{tabular}{|c||c|c|c|c|c|c|c|c|}
 \hline \rule[-2mm]{0mm}{7mm}  
 $a$ & 0 & 1 & 2 & 3 & 4 & 5 & 6 & 7 \\ \hline
 $(n,l)$ & (0,0) & $(2,-1)$ & $(-2,1)$ & $(1,0)$ & $(-1,0)$ & $(1,-1)$ & $(-1,1)$ & 
 $(3,-1)$ \\ \hline 
 $\tl{\lmd}_a$ & 0 & 3.20 & 3.20 & 4.10 & 4.10 & 4.69 & 4.69 & 5.68 
 \\\hline\hline 
 $a$ & 8 & 9 & 10 & 11 & 12 & 13 & 14 & $\cdots$ \\ \hline
 $(n,l)$ & $(-3,1)$ & $(4,-2)$ & $(-4,2)$ & $(3,-2)$ & $(-3,2)$ & (2,0) & (0,1) 
 & $\cdots$ \\ \hline
 $\tl{\lmd}_a$ & 5.68 & 6.41 & 6.41 & 6.90 & 6.90 & 8.21 & 8.21 & $\cdots$ \\ \hline
\end{tabular}
\end{center}
\caption{Relabeling the KK modes in the case of $\tau=2\exp(\pi i/8)$}
\label{relabelKK:2}
\end{table}
The number of the KK excited modes below $\Lmd$, \ie, $N_\Lmd$, grows as
\be
 N_\Lmd \propto \Lmd^2, 
\ee
except for regions: $\arg\tau\simeq 0,\pi$, $\abs{\tau}\ll 1$ and $\abs{\tau}\gg 1$, 
in which the spacetime approaches 5D and thus $N_\Lmd\propto\Lmd$. 

Then, (\ref{KKexpand}) is rewritten as
\be
 \phi(x^\mu,z) = \sum_{a=0}^\infty f_a(z)\phi_a(x^\mu). 
 \label{KKexpand2}
\ee
Plugging (\ref{KKexpand2}) into (\ref{cL2}) and performing the $d^2z$-integral, 
we obtain the 4D Lagrangian: 
\be
 \cL^{\rm (4D)} = -\sum_a\der^\mu\phi_a^*\der_\mu\phi_a
 -\sum_{a,b}M_{ab}^2\phi_a^*\phi_b+\cdots, 
\ee
where
\bea
 M_{ab}^2 \defa \frac{\tl{\lmd}_a^2}{\pi^2R^2}\dlt_{ab}
 +\cm^2f_a^*(0)f_b(0) \nonumber\\
 \eql \tl{m}_a^2\dlt_{ab}+\frac{\cm^2}{4\pi^2R^2\Im\tau} 
 \label{def:M_ab}
\eea
is the mass matrix of our theory.

\section{Cutoff dependence} \label{MassMatrix}
Since the theory is valid below $\Lmd$, we only consider 
the KK modes~$\phi_a$ ($a=0,1,\cdots,N_\Lmd$). 
Then, the mass squared eigenvalues, which are denoted as 
$\brc{m_0^2,m_1^2,\cdots,m_{N_\Lmd}^2}$, are obtained as 
eigenvalues of the finite matrix~$M_{ab}^2$ ($a,b=0,1,\cdots,N_\Lmd$). 

Since $\tl{m}_{n,l}^2=\tl{m}_{-n,-l}^2$, all the nonzero modes have degenerate modes 
when the brane mass is absent. 
Especially, $\tl{m}_1^2=\tl{m}_2^2$. 
This means that $M_{ab}^2$ has the eigenvalue~$\tl{m}_1^2$ 
with the eigenvector~$(0,1,-1,0,0,\cdots,0)$. 
In fact, this is the second smallest eigenvalue of $\tl{M}_{ab}^2$. 
Namely, the mass of the first KK excited mode~$m_1$ is independent of $\cm$ 
and $\Lmd$: 
\be
 m_1 = \tl{m}_1 = \frac{1}{R\Im\tau}\cdot\min_{(n,l)\neq (0,0)}\abs{n+l\tau}. 
 \label{expr:m1}
\ee
Thus we take $m_1$ as the compactification scale throughout the paper. 

\begin{figure}[th]
\begin{center}
\includegraphics[width=7.5cm]{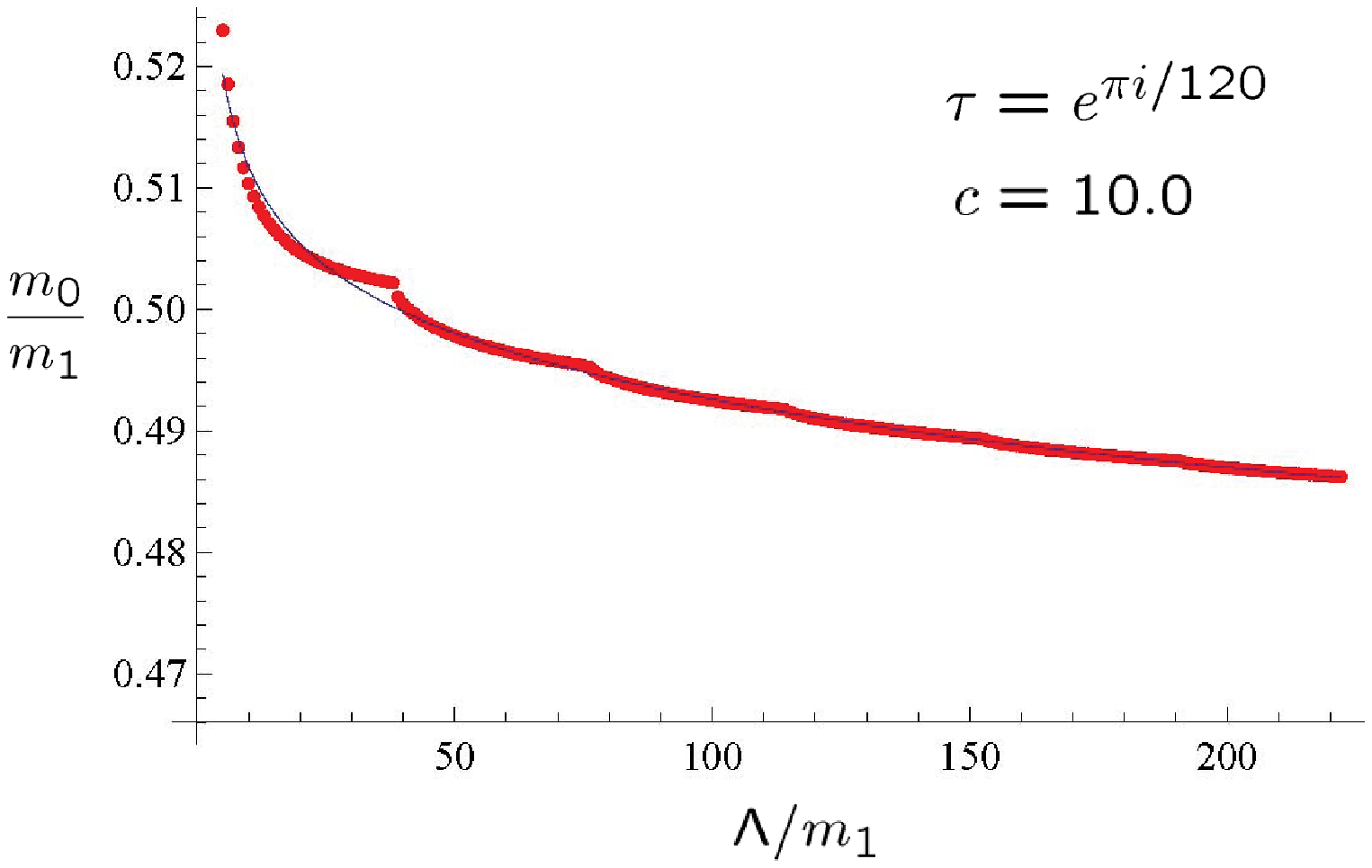}
\includegraphics[width=7.5cm]{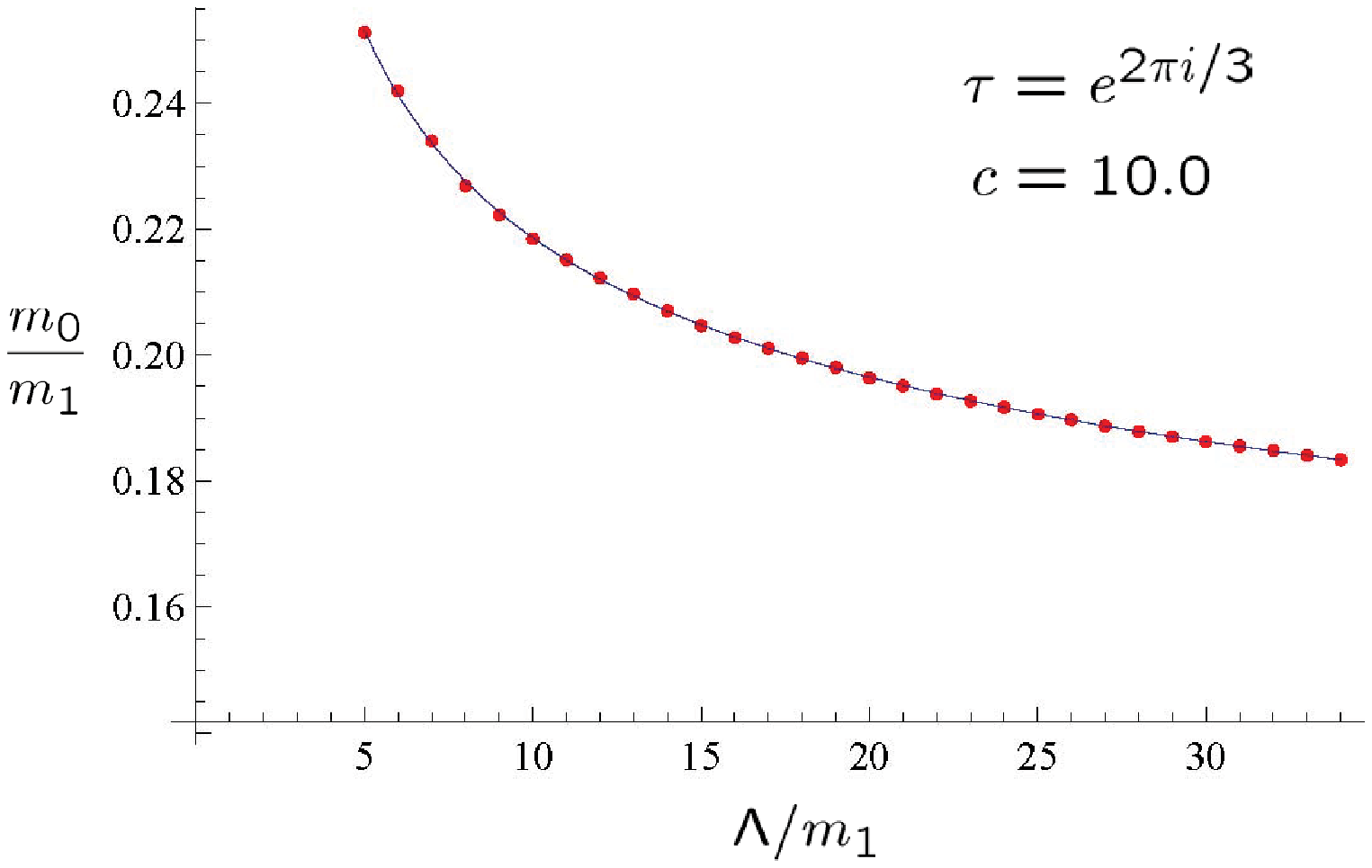}
\includegraphics[width=7.5cm]{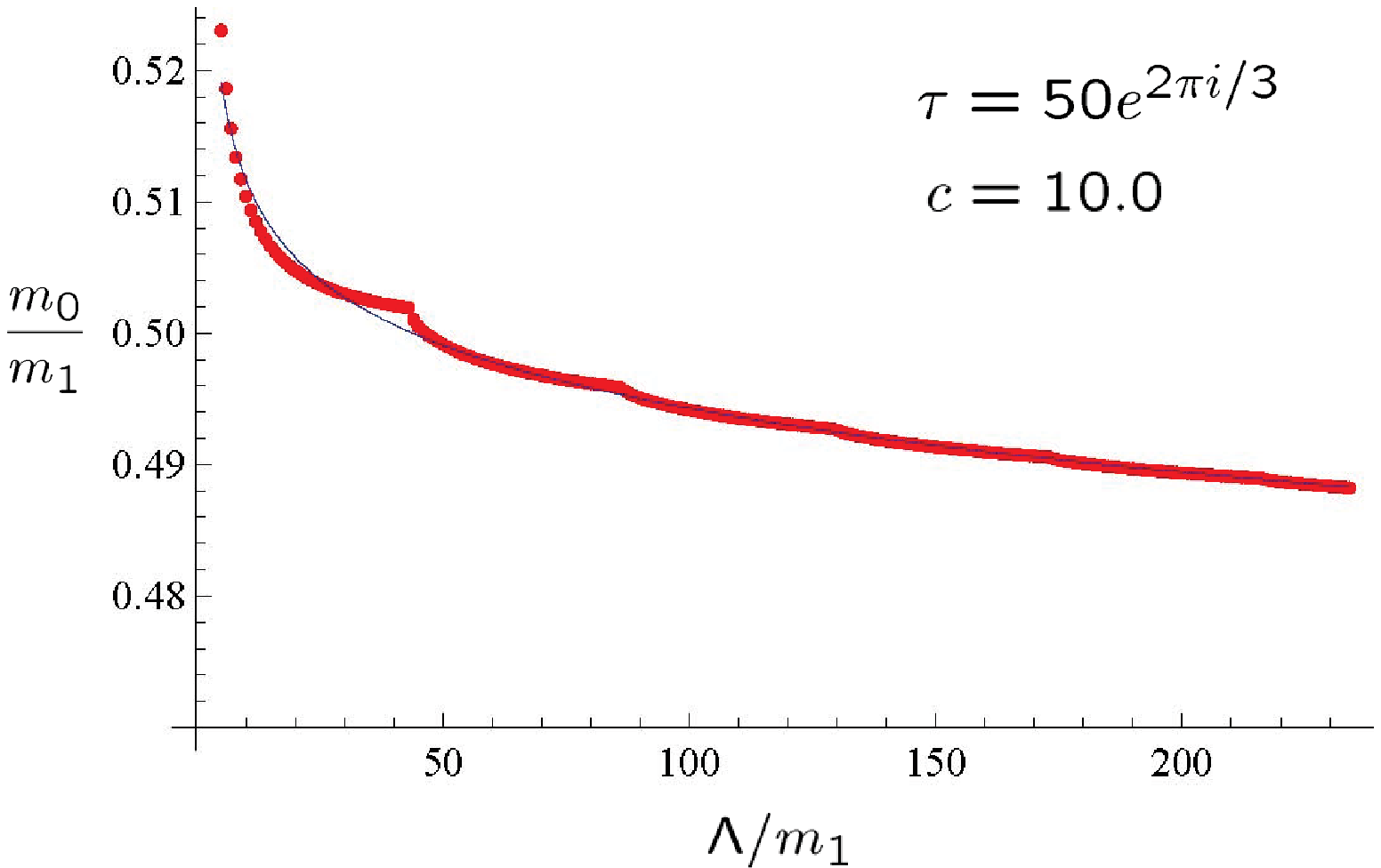}
\end{center}
\caption{The lightest mass eigenvalue~$m_0$ as a function of $\Lmd$ 
in the case of $\tau=\exp\brkt{\pi i/120}$, $\exp\brkt{2\pi i/3}$ 
and $50\exp\brkt{2\pi i/3}$ and $\cm=10.0$. 
The solid lines represent the function~(\ref{expr:lmd0}) 
with the parameters~$(\alp_1,\alp_2,\alp_3,\alp_4)=(11.9,4.01,0.0728,0.466)$,
$(1.82,3.69,0.270,0.142)$ and 
$(12.4,4.72,0.0701,0.470)$, respectively. }
\label{Lmd-dep}
\end{figure}

Plots in Fig.~\ref{Lmd-dep} show the $\Lmd$-dependence of the lightest eigenvalue~$m_0$ 
in the cases of $\tau=\exp(\frac{\pi i}{120}),\exp(\frac{2\pi i}{3})$, 
and $50\exp(\frac{2\pi i}{3})$ and $\cm=10.0$, in the unit of $m_1$. 
The right end of the horizontal axis in each plot corresponds to 
the value of $\Lmd$ such that $N_\Lmd\simeq 4000$. 
For a given value of $\cm$, the ratio~$m_0/m_1$ can be approximated by  
\be
 \frac{m_0}{m_1} \simeq \brkt{\alp_1+\alp_2\ln\frac{\Lmd}{m_1}
 +\alp_3\frac{\Lmd}{m_1}}^{-1}+\alp_4, 
 \label{expr:lmd0}
\ee
where $\alp_i$ ($i=1,2,3,4$) are real constants. 
The solid lines in Fig.~\ref{Lmd-dep} represent the fitting functions 
of the form~(\ref{expr:lmd0}). 
The constant~$\alp_4$ is the asymptotic value of $m_0/m_1$ in the limit of $\Lmd\to\infty$:
\be
 \lim_{\Lmd\to\infty}\frac{m_0(\Lmd)}{m_1} = \alp_4. \label{alp4}
\ee
The horizontal axes in Fig.~\ref{Lmd-dep} denote the asymptotic lines 
that the curves approach. 
Typically, $m_0$ approaches to the limit value 
much more slowly compared with the 5D case 
(see Fig.~\ref{Lmd-dep:5D} in Appendix~\ref{5D:num}).
Thus the cutoff dependence of the spectrum cannot be neglected 
even when $\Lmd/m_1=\cO(100)$. 
This cutoff dependence becomes smaller 
when $\arg\tau\simeq 0,\pi$ or $\abs{\tau}\ll 1$ or $\abs{\tau}\gg 1$. 
This is because the torus is squashed or stretched in such cases, 
and the spacetime approaches to 5D. 
In fact, as we can see from Fig.~\ref{Lmd-dep}, 
\be
 \frac{m_0(15m_1)}{m_1} \simeq \lim_{\Lmd\to\infty}\frac{m_0(\Lmd)}{m_1} \times
 \begin{cases} 1.40 & (\tau=e^{2\pi i/3}) \\ 1.07 & 
(\tau=e^{\pi i/120},\; 50e^{2\pi i/3}) 
 \end{cases}. 
\ee
Note that the curve for $\Lmd<40m_1$ in the top-left plot 
or in the bottom plot are almost the same 
as that of the 5D case (shown in Fig.~\ref{Lmd-dep:5D}). 
The cusp at $\Lmd=40m_1$ indicates that the field begins to feel the width of 
the squashed torus or the smaller cycle of the long thin torus. 

\begin{figure}[t]
\begin{center}
\includegraphics[width=8cm]{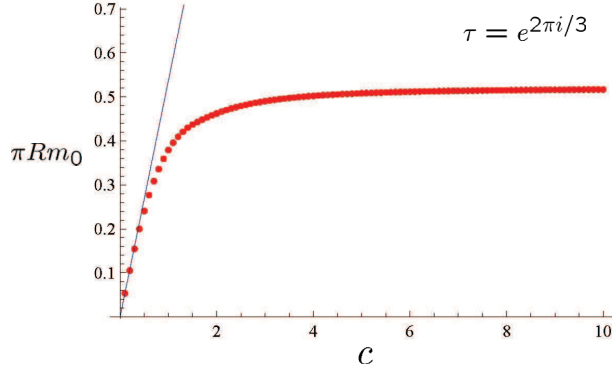}
\end{center}
\caption{The lightest mass eigenvalue~$m_0$ as a function of $\cm$ 
in the case of $\tau=e^{2\pi i/3}$. 
The solid line represents (\ref{lmd0:smallc}). }
\label{vsc}
\end{figure}

In the following, we focus on the limit value~(\ref{alp4}). 
Fig.~\ref{vsc} shows its dependence on the brane mass~$\cm$. 
The unit here is taken as $1/(\pi R)$. 
For small values of $\cm$, the lightest mass eigenvalue~$m_0$ 
is approximated as 
\be
 m_0 \simeq \sqrt{M_{00}^2} = \frac{\cm}{2\pi R\sqrt{\Im\tau}}, 
 \label{lmd0:smallc}
\ee
which is plotted as the solid line in Fig.~\ref{vsc}. 
This is because the brane mass can be treated as a perturbation in this region, 
and the mixing among the KK modes induced by it is negligible. 
As the brane mass grows, such mixing effect becomes significant, 
and $m_0$ saturates and is almost independent of $\cm$ when $\cm\simgt 5$. 
This situation is the same as the 5D case (see Fig.~\ref{vsc:5D} in Appendix~\ref{5D:anl}). 
In the following discussion, we take $c=10.0$ as a representative of $\cm\gg 1$.

\section{Approximate expression} \label{ap_expr}

\begin{figure}[t]
\begin{center}
\includegraphics[width=7cm]{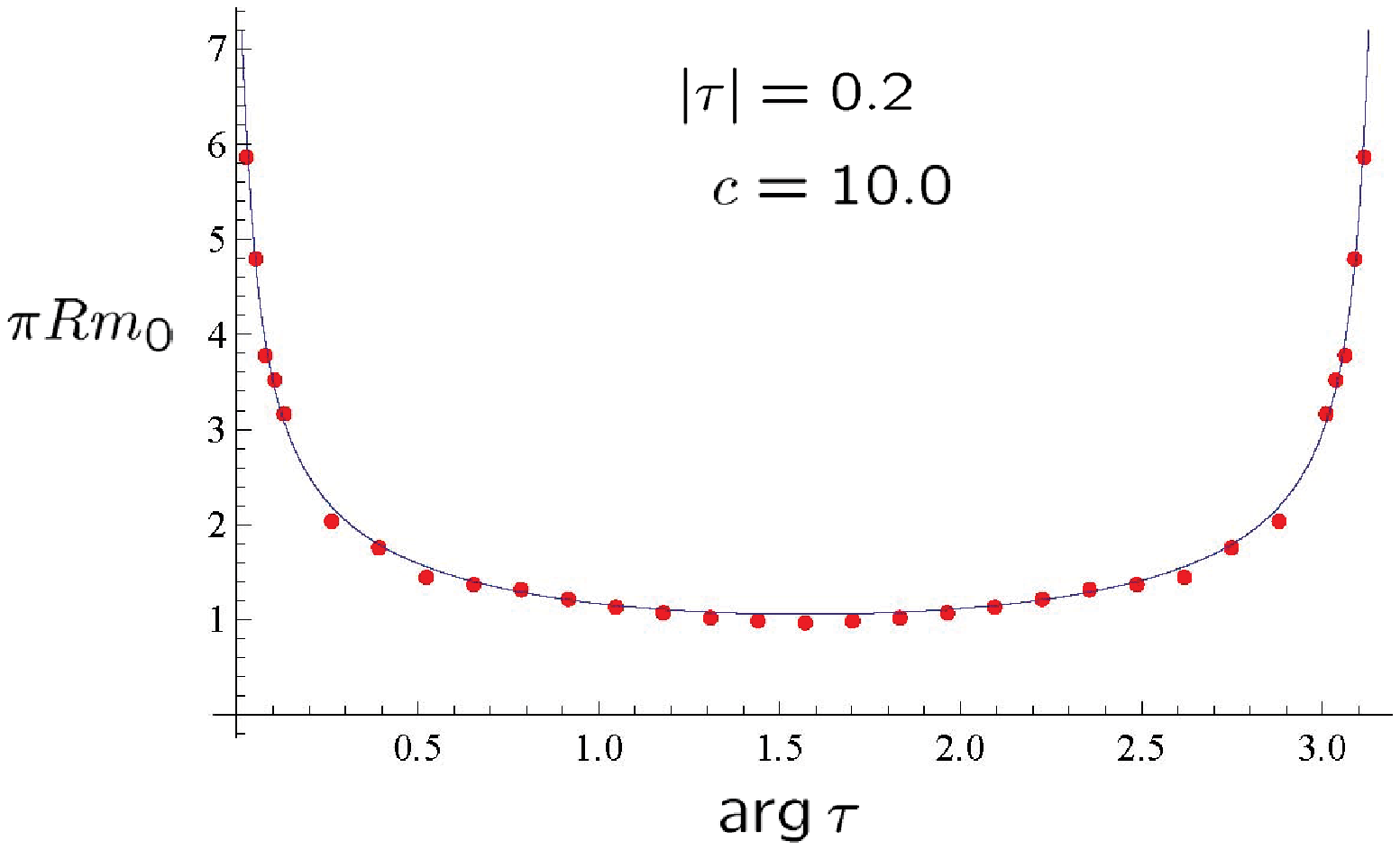}
\includegraphics[width=7cm]{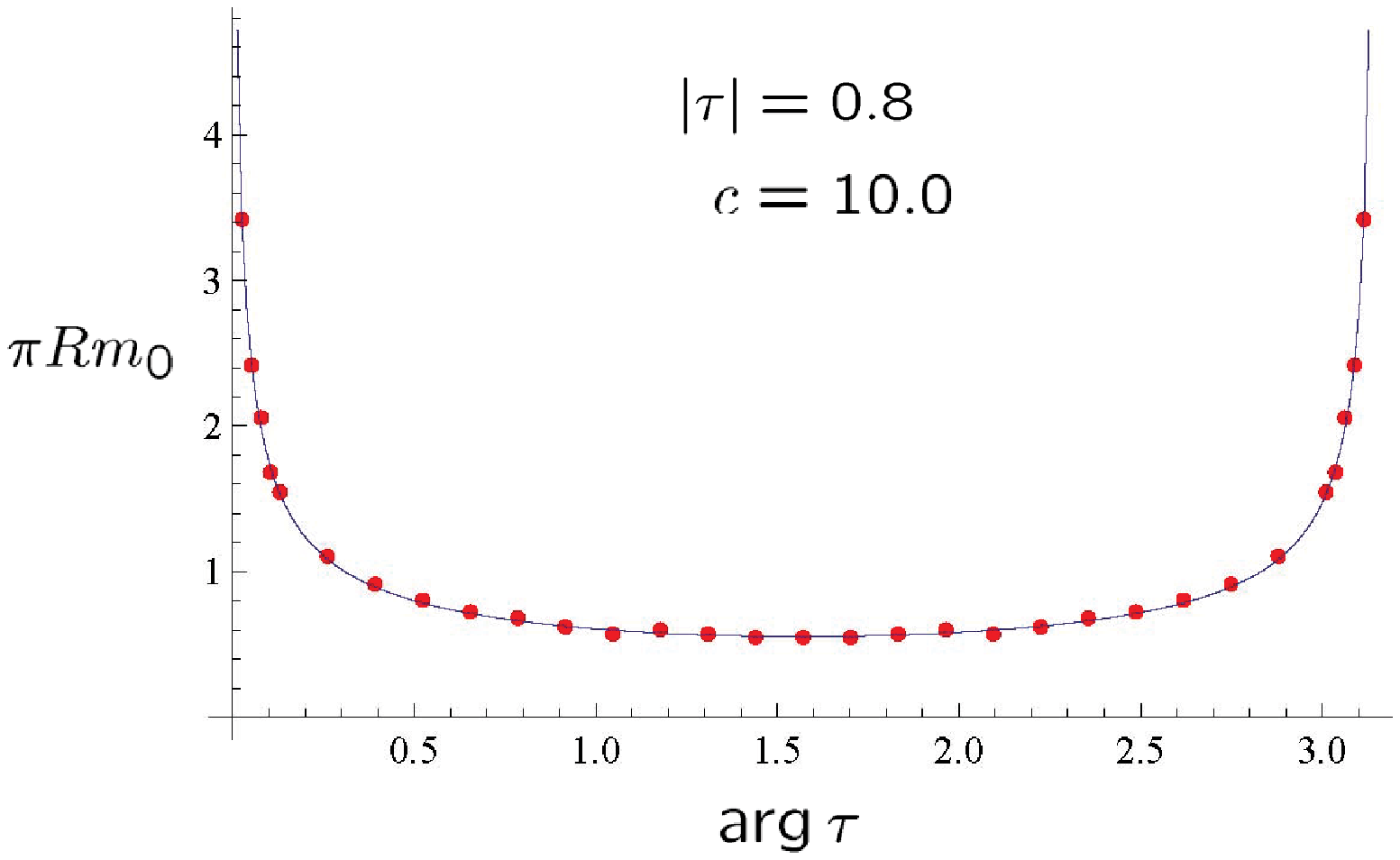}
\includegraphics[width=7cm]{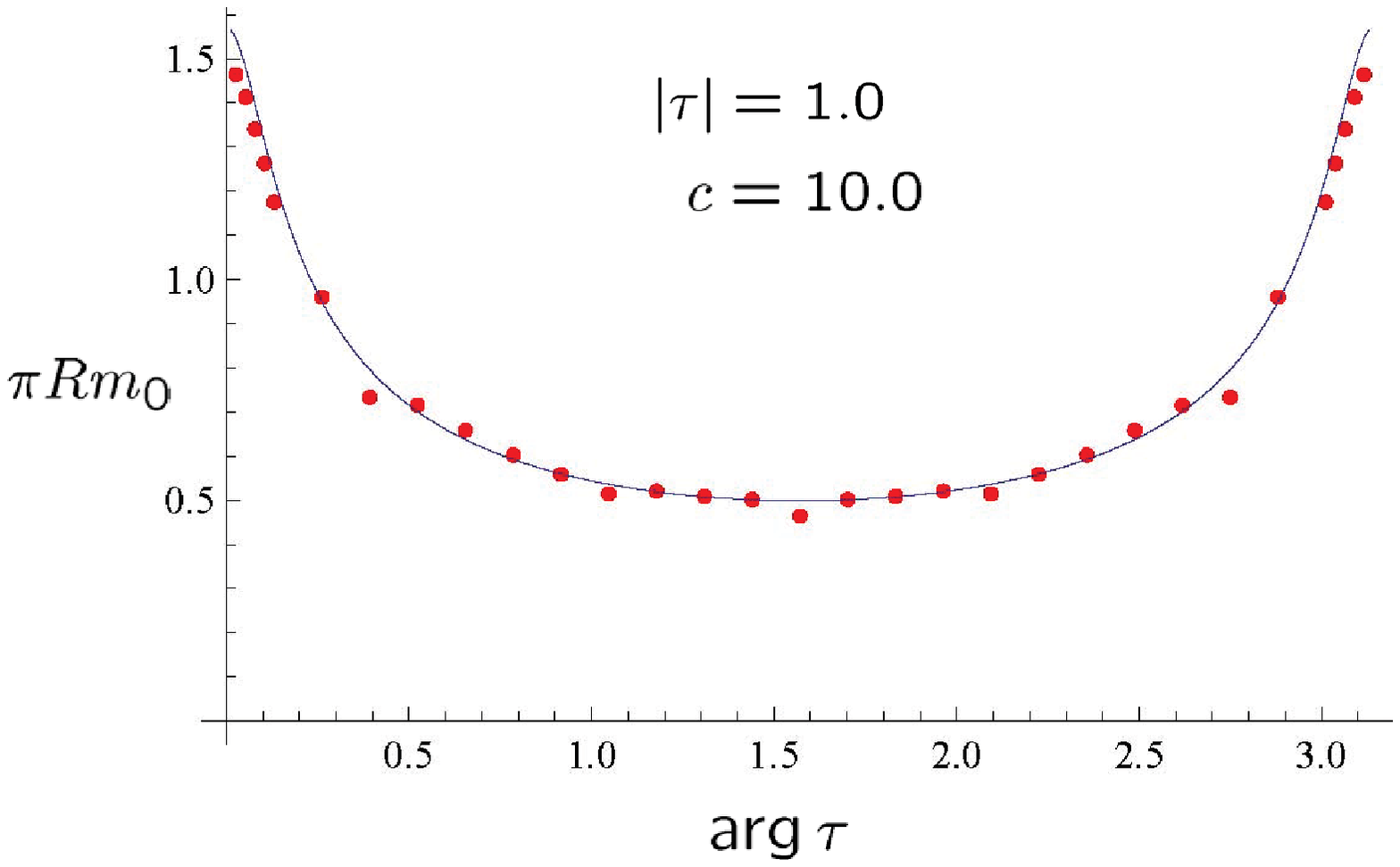}
\includegraphics[width=7cm]{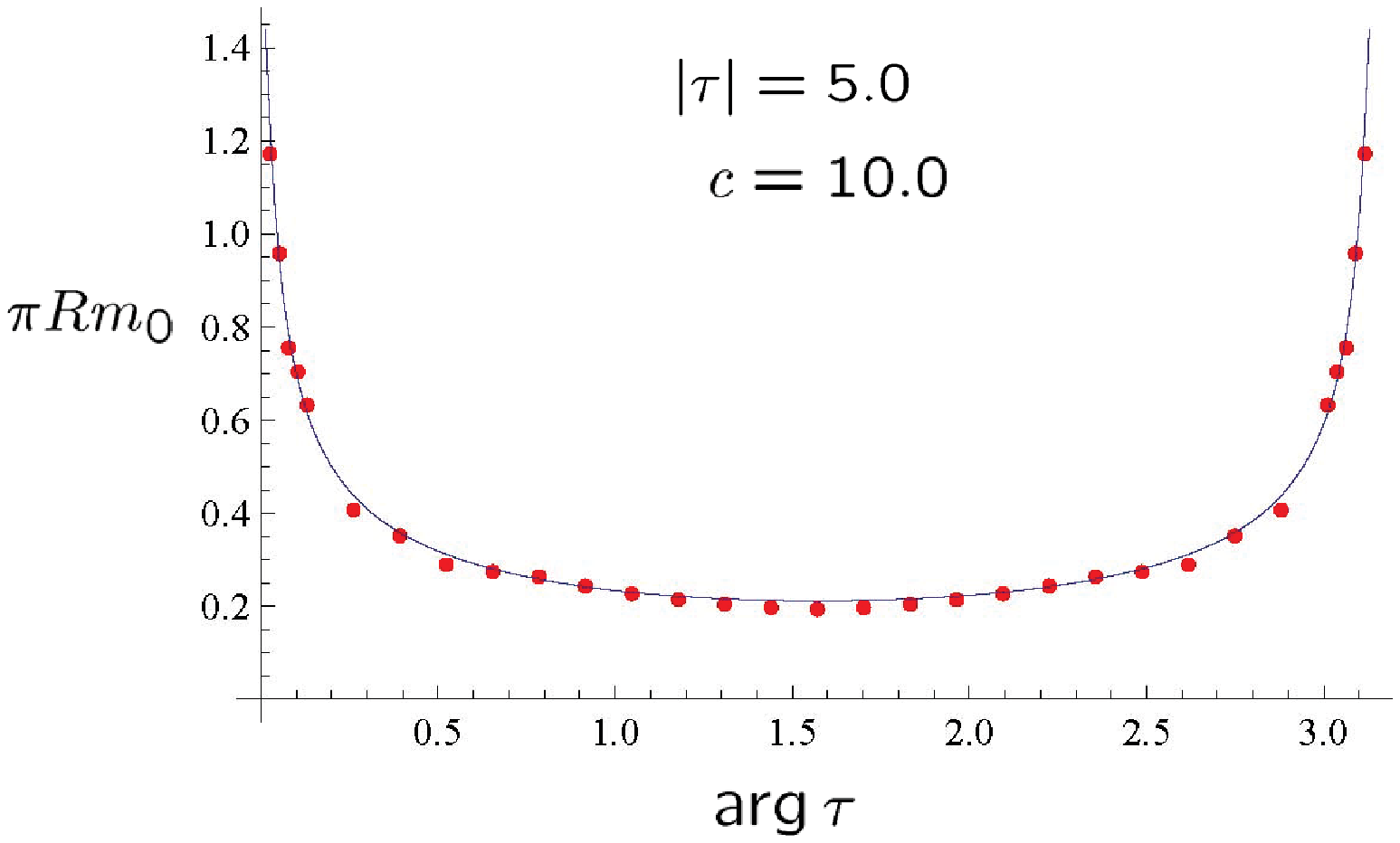}
\end{center}
\caption{The lightest mass eigenvalue~$m_0$ as a function of $\arg\tau$ 
for various values of $\abs{\tau}$. 
The solid lines represent the approximate expression~(\ref{ap:m_0}). }
\label{lmdvstht}
\end{figure}
Fig.~\ref{lmdvstht} shows the dependence of $m_0$ on $\arg\tau$ 
for various values of $\abs{\tau}$. 
Here we will find an approximate expression of $m_0$ as a function of $\tau$. 

First, we should note that the mass eigenvalues~$m_a$ are 
functions of $\cm$ and $\tau$, and should satisfy 
\be
 m_a\brkt{\cm;-\frac{1}{\tau}} = \abs{\tau}m_a(\cm;\tau), 
 \label{cond1_for_lmd0}
\ee
since the theory is defined on the torus. 
Besides, from (\ref{wocm}) and (\ref{def:M_ab}), we also find that
\be
 m_a(\cm;-\bar{\tau}) = m_a(\cm;\tau). 
 \label{cond2_for_lmd0}
\ee

As mentioned in the previous section, 
there are two limits in which the spacetime approaches to 5D, \ie, 
$\arg\tau\to 0,\pi$ (squashed torus) and $\abs{\tau}\to 0,\infty$ (stretched torus). 
In these cases, the low-lying KK masses in the absence of the brane mass 
are approximately expressed as follows. 
\begin{description}
\item[$\bdm{\abs{\tau}\gg 1}$] 
\be
 \tl{m}_a = \tl{m}_{n(a),0} 
 \simeq \frac{\abs{n(a)}}{R\Im\tau},  \label{apex1:ma}
\ee
where $a\simlt 2\abs{\tau}$, 
and $n(a)\equiv (-1)^a{\rm floor}\brkt{\frac{a+1}{2}}$. 

\item[$\bdm{\tht\equiv\arg\tau\ll 0}$] 
\be
 m_{n,l} = \frac{1}{R}\brc{\frac{(n+l\abs{\tau}\cos\tht)^2}
 {\abs{\tau}^2\sin^2\tht}+l^2}^{1/2}
 \simeq \frac{1}{R}\brc{\frac{1}{\tht^2}\brkt{\frac{n}{\abs{\tau}}+l}^2+l^2}^{1/2}. 
\ee
Especially when $\abs{\tau}$ is a rational number, \ie, $\abs{\tau}=p/q$ 
($p$ and $q$ are relatively prime integers and $q>0$), 
the light masses are approximated as 
\be
 m_a = m_{n(a)p,-n(a)q} \simeq \frac{\abs{n(a)q}}{R}, 
 \label{apex2:ma}
\ee
where $a\simlt\frac{2}{\tht}\min(1,\abs{\tau^{-1}\pm 1})$. 
\end{description}
As for the cases of $\abs{\tau}\ll 1$ and of $\pi-\arg\tau\ll 1$, 
approximate expressions of $m_a$ are obtained from (\ref{apex1:ma}) and (\ref{apex2:ma}) 
by using (\ref{cond1_for_lmd0}) and (\ref{cond2_for_lmd0}), respectively. 
Then, we identify the effective radius of $S^1$ as
\be
 R_{\rm eff} = \begin{cases} R\Im\tau & (\abs{\tau} \gg 1) \\
 R\Im\tau/\abs{\tau} & (\abs{\tau} \ll 1) \\
 R/q & (\arg\tau \ll 1 \;\;\mbox{or}\;\; \pi-\arg\tau\ll 1) \end{cases}. 
\ee
Using this, the low-lying KK masses~$m_a$ can be expressed as
(see Appendix~\ref{5D:anl}) 
\be
 m_a \simeq \frac{\abs{n(a)}}{R_{\rm eff}}, 
\ee
or solutions of 
\be
 m_a \simeq \frac{\hat{c}_{\rm eff}^2}{2}\cot(\pi R_{\rm eff}m_a), 
 \label{cond3_for_lmd0}
\ee
where the ``effective 5D brane mass''~$\hat{c}_{\rm eff}$ is defined as 
\be
 \hat{c}_{\rm eff}^2 \equiv \frac{R_{\rm eff}\cm^2}{2\pi R^2\Im\tau}, 
\ee
which is identified from the condition that (\ref{lmd0:smallc}) is reproduced. 
When $c$ is sufficiently large, the solutions of (\ref{cond3_for_lmd0}) are
\be
 m_a \simeq \frac{\abs{n(a)+\frac{1}{2}}}{R_{\rm eff}}. 
\ee
Especially, the lightest mass eigenvalue is
\be
 m_0 \simeq \frac{1}{2R_{\rm eff}} \simeq \frac{m_1}{2}. 
 \label{cond4_for_lmd0}
\ee

Taking into account the properties~(\ref{cond1_for_lmd0}), 
(\ref{cond2_for_lmd0}) and (\ref{cond4_for_lmd0}), 
we find an approximate expression of $m_0$ that fits Fig.~\ref{lmdvstht} as
\be
 m_0^{\rm (ap)} =  \frac{\sqrt{\sin\brc{\arcsin(\tl{\lmd}_1^2\Im\tau)}}}
 {2\pi R\sqrt{\Im\tau}}.  \label{ap:m_0}
\ee
This is plotted as solid lines in Fig.~\ref{lmdvstht}. 

\begin{figure}[t]
\begin{center}
\includegraphics[width=7cm]{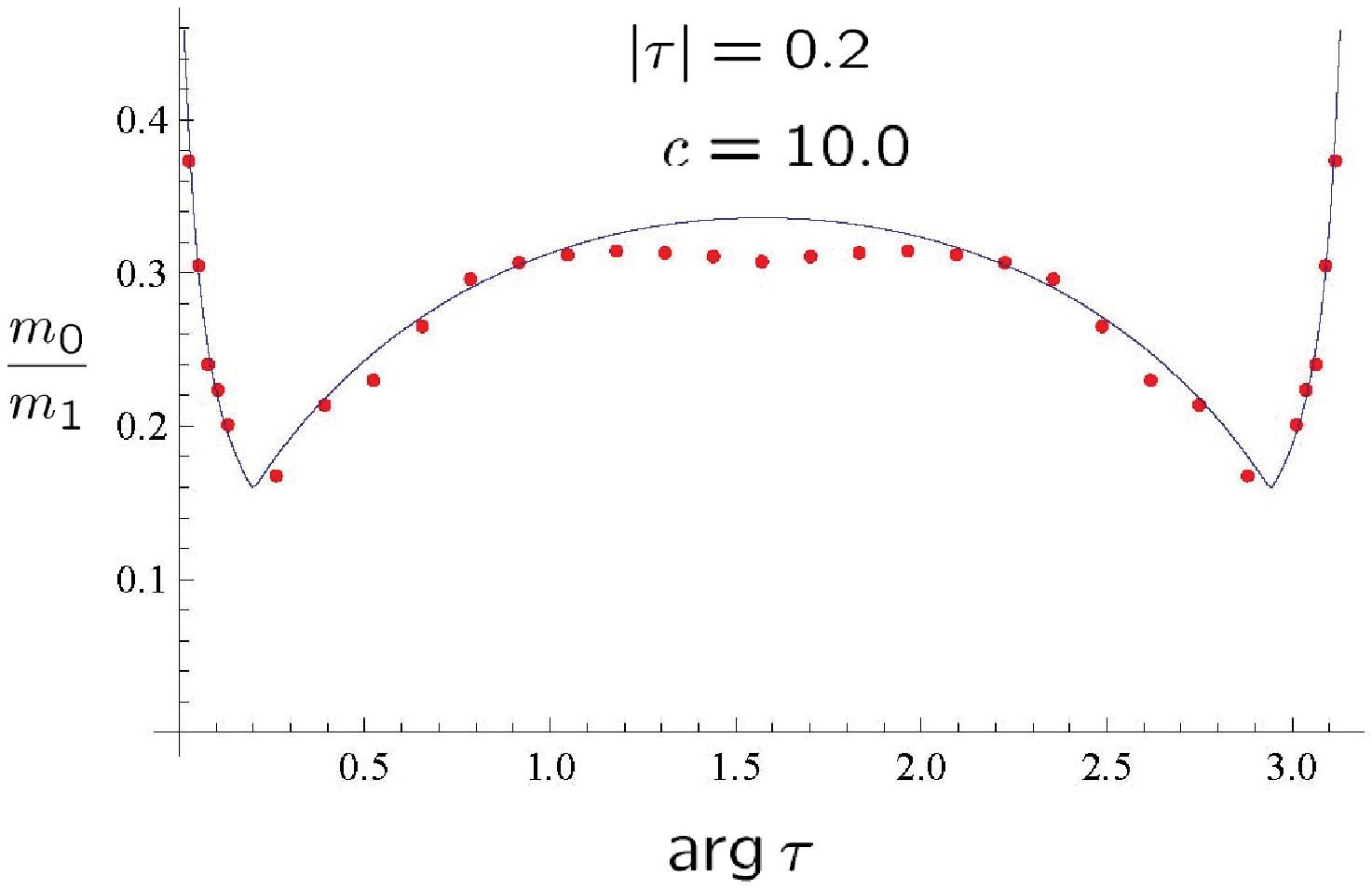}
\includegraphics[width=7cm]{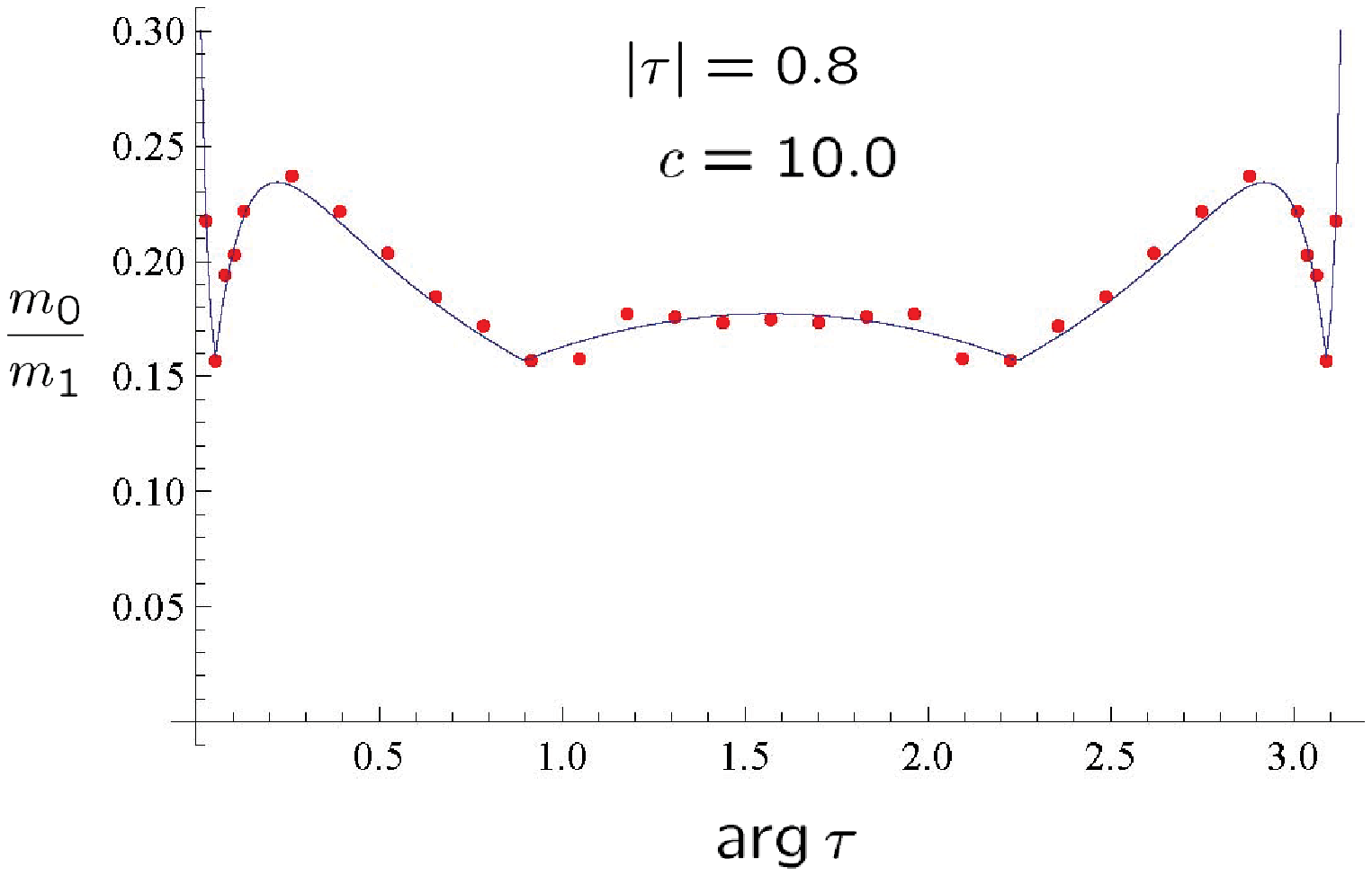}
\includegraphics[width=7cm]{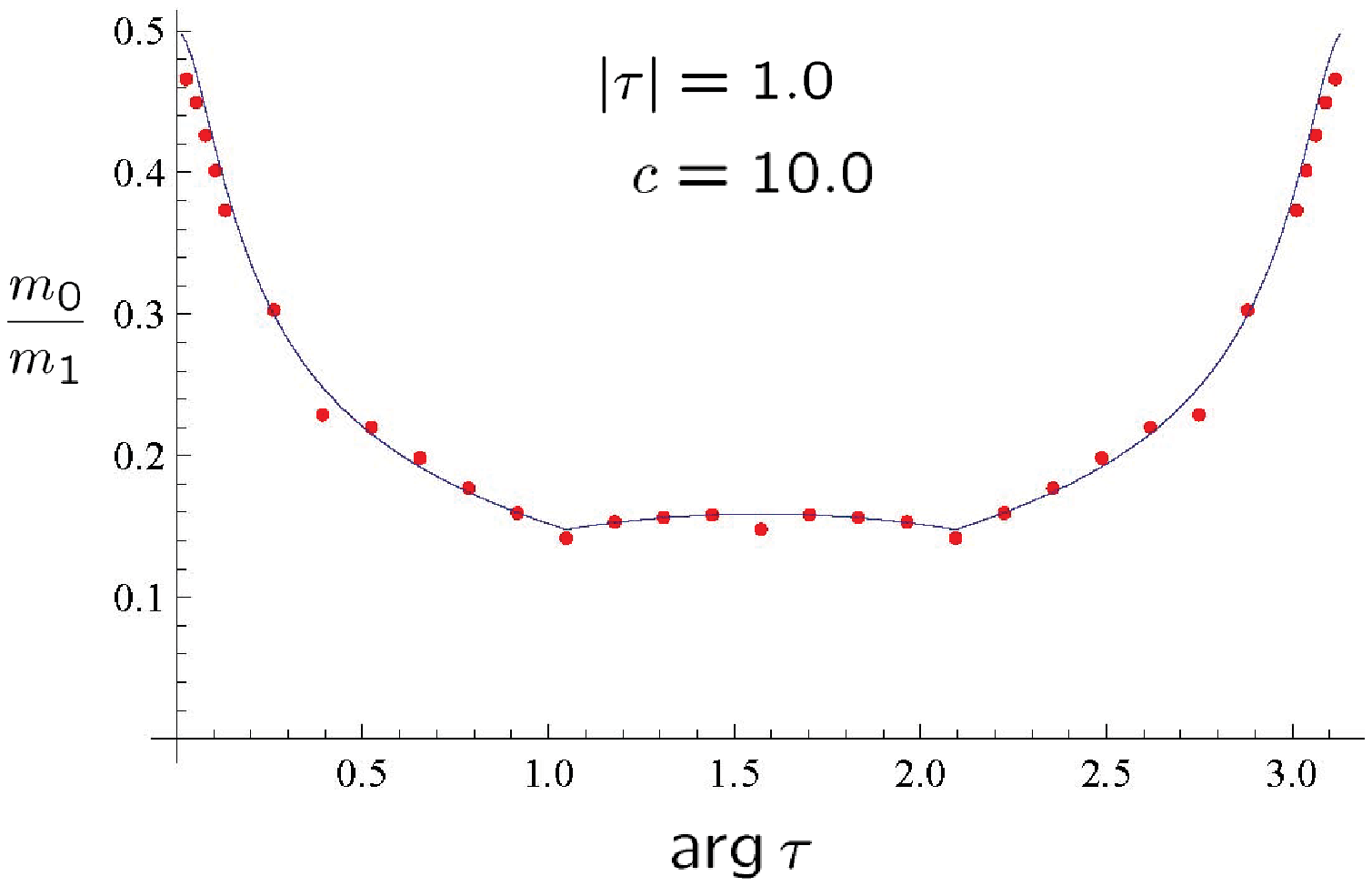}
\end{center}
\caption{The ratio of the lightest mass eigenvalue~$m_0$ 
to the compactification~$m_1$. 
The solid lines represents the ratio of (\ref{ap:m_0}) to (\ref{expr:m1}). }
\label{rtovstht}
\end{figure}

Finally, we evaluate the ratio of $m_0$ 
to the compactification scale~$m_1$. 
As Fig.~\ref{rtovstht} shows, this ratio is much smaller than the value 
of the 5D case, 1/2, except for the extreme cases 
in which the spacetime is 5D-like. 
Typically, $m_0$ is lighter than $m_1$ by one order of magnitude. 
Namely, the brane-localized mass cannot make the zero-modes 
as heavy as the compactification scale. 
This is an important fact in model building.

\section{Summary and comments} \label{summary}
\subsection{Summary}
We have evaluated the mass eigenvalues of a 6D theory compactified on a torus 
in the presence of the brane-localized mass term. 
Especially we focus on the lightest mode that becomes massless  
in the zero brane-mass limit. 

From the numerical calculations, 
we confirmed that the lightest mass eigenvalue~$m_0$ 
has non-negligible dependence on the cutoff scale~$\Lmd$ 
even when $\Lmd$ is larger than the compactification scale 
by two orders of magnitude. 
This indicates that $m_0$ is sensitive to the internal structure of 
the brane when the brane has a finite size. 
This is consistent with the results in Ref.~\cite{Dudas:2005vn}. 

We find an approximate expression of $m_0$ which is valid for a large brane mass. 
It clarifies the dependence on the size and the shape of the torus, 
and reduces to the known result in the 5D case when the torus is squashed or 
stretched. 

In contrast to the 5D case, $m_0$ is much smaller than 
the compactification scale unless the torus is squashed or stretched. 
Their ratio is typically $\cO(0.1)$. 
This is because the effects of the brane term are spread out 
over the codimension two compact space and diluted. 
Hence we should be careful in model building 
especially when we introduce the brane mass terms 
in order to decouple unwanted modes. 

Although we have not discussed in this paper, 
the brane mass also deforms the profiles of the mode functions. 
They can be obtained by calculating the eigenvectors of $M_{ab}^2$ 
in (\ref{def:M_ab}). 
The main effect of the brane mass on the mode functions is 
to push them out from the position of the brane. 
Namely, it reduces their absolute values at the brane to zero.

\subsection{On more general setups}
We have discussed in a theory of a scalar field because it is the simplest case. 
However, the properties of the spectrum clarified in the text are also found 
in cases of fermion and vector fields, as shown in Appendix~\ref{other_cases}. 
So our result is valid in a wider class of 6D theory. 

Besides, we have assumed that the bulk mass is zero and the brane squared mass 
is positive. 
In the presence of the bulk mass~$M_{\rm bk}$, 
the mass matrix~(\ref{def:M_ab}) becomes 
\be
 M_{ab}^2 = \brkt{M_{\rm bk}^2+\tl{m}_a^2}\dlt_{ab}+\frac{c^2}{4\pi^2R^2\Im\tau}, 
\ee
where $\tl{m}_{n,l}=\abs{n+l\tau}/(R\Im\tau)$. 
Thus the bulk mass just raises the whole spectrum. 
However, if we allow a tachyonic brane mass, \ie, $c^2<0$, 
a light mode may appear below the compactification scale.   
If $\abs{c}^2$ is large enough, $m_0$ becomes tachyonic 
and thus $\vev{\phi}=0$ is no longer the vacuum. 
In such a case, $\phi$ has a nontrivial background that depend on 
the extra-dimensional coordinates~$z$ and $\bar{z}$, and we have to expand $\phi$ 
around it in order to obtain the mass matrix~$M_{ab}^2$. 
It is not an easy work to find such a nontrivial background.  
Here we do not discuss this issue further, but give a comment on it. 
Note that the smallest diagonal 
element~$M_{00}^2=M_{\rm bk}^2+c^2/(4\pi^2R^2\Im\tau)$ provides the upper bound on $m_0$. 
Thus, $2\pi R\sqrt{\Im\tau}M_{\rm bk}>\abs{c}$ must be satisfied 
in order to avoid the vacuum instability for $\vev{\phi}=0$. 
In other words, there is a value of $c$ that leads to a tachyonic mass eigenvalue 
no matter how large $M_{\rm bk}$ is. 
This indicates that the effect of the brane mass on the spectrum does not saturate,  
which is in contrast to the non-tachyonic brane mass. 
The mode function is attracted toward the brane by the tachyonic brane mass. 

We considered the scalar field with the periodic boundary condition. 
Twisted boundary conditions are also allowed, but they just raise the mass spectrum. 
This can be understood from the fact that imposing the twisted boundary conditions is 
equivalent to introducing a non-vanishing background gauge field coupled to 
the scalar field with the periodic boundary conditions. 
Such a background gauge field play the same role as the bulk scalar mass $M_{\rm bk}$ 
mentioned above. 

We have also assumed that the spacetime is flat, 
no background magnetic fluxes exist,\footnote{
The introduction of the background fluxes leads to the multiplication 
of the modes at each KK level. 
Thus the size of the mass matrix~(\ref{def:M_ab}) becomes larger, 
and it will take much more time to calculate the mass eigenvalues. 
So we need to develop more efficient way to discuss in such a case. 
}
and there is only one brane, for simplicity. 
It is an interesting and useful extension to relax these assumptions. 
This will be discussed in separate papers.

\subsection*{Acknowledgements}
The author would like to thank Yukihiro Fujimoto for valuable information. 
This work was supported in part by 
Grant-in-Aid for Scientific Research (C) No.~25400283  
from Japan Society for the Promotion of Science (Y.S.).

\appendix

\section{5D case} \label{5Dcase}
Here we summarize the effects of the brane-localized mass 
in a 5D complex scalar theory. 
The Lagrangian is
\be 
 \cL = -\der^{\hat{M}}\phi^*\der_{\hat{M}}\phi
 -\hcm^2\abs{\phi}^2\dlt(x^4)+\cdots,  \label{cL:5D}
\ee
where $\hat{M}=0,1,2,3,4$, and the ellipsis denotes interaction terms. 
The brane mass parameter~$\hcm$ is a real dimension 1/2 constant. 
The extra dimension is compactified on $S^1$ whose radius is $R$.

\subsection{Analytic expressions} \label{5D:anl}
The KK expansion of $\phi$ is
\be
 \phi(x^\mu,x^4) = \sum_{n=-\infty}^\infty f_n(x^4)\phi_n(x^\mu). 
\ee
The mode function~$f_n(x^4)$ satisfies the mode equation, 
which is read off from (\ref{cL:5D}) as 
\be
 \brc{\der_4^2-\hcm^2\dlt(x^4)}f_n(x^4) = -m_n^2f_n(x^4). 
 \label{md_eq:5D}
\ee
By integrating this over an infinitesimal interval~$[-\ep,\ep]$, we obtain
\be
 \sbk{\der_4f_n}_{-\ep}^\ep-\hcm^2 f_n(0) = 0.  \label{BC:5D}
\ee
Thus the brane mass changes the boundary condition of the bulk field. 

In the bulk region~$[\ep,2\pi R-\ep]$, (\ref{md_eq:5D}) is solved as 
\be
 f_n(x^4) = \cC_{+n}e^{im_n x^4}+\cC_{-n}e^{-im_n x^4}. 
\ee
where $\cC_{\pm n}$ are complex constants. 
From the periodic condition~$f_n(x^4+2\pi R)=f_n(x^4)$ and (\ref{BC:5D}), 
we obtain 
\bea
 &&\cC_{+n}+\cC_{-n} = \cC_{+n}e^{2\pi im_nR}+\cC_{-n}e^{-2\pi im_nR}, \nonumber\\
 &&im_n\brkt{\cC_{+n}-\cC_{-n}}-im_n\brkt{\cC_{+n}e^{2\pi im_nR}-\cC_{-n}e^{-2\pi im_nR}}
 = \hcm^2\brkt{\cC_{+n}+\cC_{-n}}. 
\eea
The solutions of these equations are 
\be
 \cC_{-n} = -\cC_{+n}, \;\;\;\;\;
 m_n = \frac{\abs{n}}{R}, \;\;\;\;\; (n\neq 0) \label{5D:sol1}
\ee
or
\be
 \cC_{-n} = e^{2\pi im_nR}\cC_{+n}, \;\;\;\;\;
 m_n = \frac{\hcm^2}{2}\cot(\pi Rm_n).  \label{5D:sol2}
\ee
When $\hcm=0$, the spectrum determined by the second equation of (\ref{5D:sol2}) 
coincides with that of (\ref{5D:sol1}). 
Note that the mode functions corresponding to (\ref{5D:sol1}) are odd functions. 
Thus, they do not feel the brane-localized mass because they vanish at $x^4=0$. 

\begin{figure}[t]
\begin{center}
\includegraphics[width=8cm]{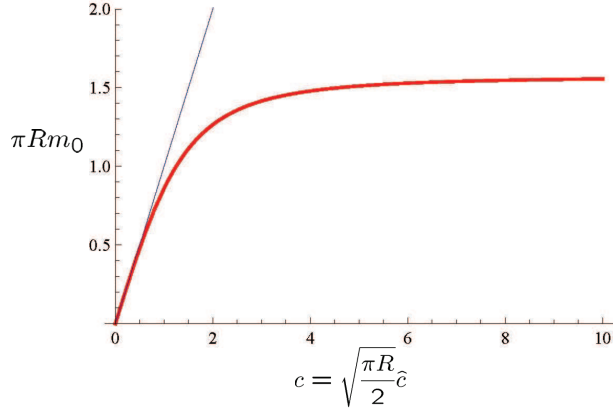}
\end{center}
\caption{The lightest mass eigenvalue~$m_0$ as a function of $\hcm$. }
\label{vsc:5D}
\end{figure}

Fig.~\ref{vsc:5D} shows the lightest mass~$m_0$ as a function of the brane mass~$\hcm$. 
For small values of the brane mass, $m_0$ is proportional to $\hcm$. 
This is because the brane mass can be treated as a perturbation in this region, 
and the mixing among the KK modes induced by it is negligible. 
As the brane mass grows, such mixing effect becomes significant, 
and $m_0$ saturates and is almost independent of $\hcm$ 
when $c\equiv\sqrt{\pi R/2}\hcm\simgt 5$. 
In the limit of $\hcm\to\infty$, the spectrum determined by (\ref{5D:sol2}) is 
\be
 m_n = \frac{\abs{n+\frac{1}{2}}}{R}.  \label{m_n:hcminf}
\ee
Since the second smallest solution of the second equation in (\ref{5D:sol2}) 
are greater than $1/R$, the first KK excited mass is $m_1=1/R$, 
which is independent of $\hcm$ and taken as the compactification scale.

\subsection{Numerical evaluation} \label{5D:num}
In order to see the cutoff dependence of the spectrum 
and compare it with that in the 6D case, 
we follow the same procedure as in Sec.~\ref{MassMatrix}. 

We relabel the KK modes by using the KK label~$a=0,1,2,\cdots$, which is defined as 
\be
 a = \begin{cases} 2n & (n\geq 0) \\ 2\abs{n}-1 & (n<0) \end{cases}, 
\ee
and expand $\phi$ as
\be
 \phi(x^\mu,x^4) = \sum_{a=0}^\infty \frac{e^{in(a)\pi x^4/R}}{\sqrt{2\pi R}}
 \phi_a(x^\mu), 
\ee
where $n(a)\equiv (-1)^a{\rm floor}\brkt{\frac{a+1}{2}}$. 
This is the KK expansion in the absence of the brane mass. 
Then, we can rewrite the 5D Lagrangian~(\ref{cL:5D}) in terms of $\phi_a$ as
\be
 \cL^{\rm (4D)} = -\sum_a\der^\mu\phi^*\der_\mu\phi
 -\sum_{a,b}\hat{M}_{ab}^2\phi_a^*\phi_b+\cdots, 
\ee
where
\be
 \hat{M}_{ab}^2 \equiv \brkt{\frac{n(a)}{R}}^2\dlt_{ab}
 +\frac{\hcm^2}{2\pi R}. 
\ee

We consider only the modes whose masses are below the cutoff scale~$\Lmd$. 
\be
 0 < \frac{\abs{n(N_\Lmd)}}{R} < \Lmd \leq \frac{\abs{n(N_\Lmd+1)}}{R}. 
\ee
Then the KK mass eigenvalues~$\brc{m_0^2,m_1^2,\cdots,m_{N_\Lmd}^2}$ 
are calculated as eigenvalues of 
the finite matrix~$\hat{M}_{ab}^2$ ($a,b=0,1,\cdots,N_\Lmd$). 

\begin{figure}[t]
\begin{center}
\includegraphics[width=7cm]{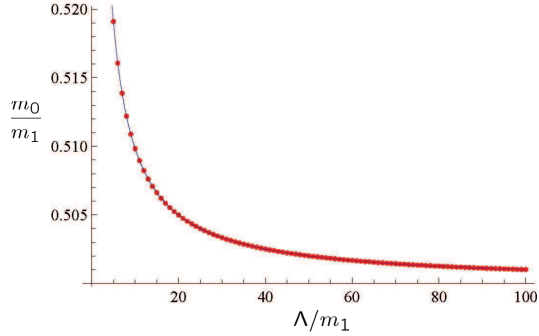}
\end{center}
\caption{The lightest mass eigenvalue~$m_0$ as a function of $\Lmd$ 
in the unit of the compactification scale~$m_1$. 
The brane mass is chosen as $c=100$. }
\label{Lmd-dep:5D}
\end{figure}

Fig.~\ref{Lmd-dep:5D} shows the lightest mass eigenvalue~$m_0$ as a function of $\Lmd$ 
in the unit of $m_1=1/R$ when $c=100$. 
The solid line represents  
\be
 \frac{m_0(\Lmd)}{m_1} = \brkt{3.01+9.87\frac{\Lmd}{m_1}}^{-1}+0.500. 
\ee
We can see from Fig.~\ref{Lmd-dep:5D} that $m_0$ rapidly approaches to $m_1/2=1/(2R)$, 
which is consistent with (\ref{m_n:hcminf}). 
The cutoff dependence is negligible when $\Lmd\simgt 10m_1$.

\section{Cases of spinor and vector fields} \label{other_cases}
\subsection{Brane mass for spinor fields}
We consider a theory which has a 6D Weyl spinor field~$\Psi_+$ 
whose 6D chirality is $+$. 
We can introduce the following brane mass term 
with the 4D spinor field localized on the brane. 
\be
 \cL = i\bar{\Psi}_+\Gm^M\der_M\Psi_+
 +\brc{-i\chi\sgm^\mu\der_\mu\bar{\chi}+c\brkt{\psi\chi+\bar{\psi}\bar{\chi}}}
 \dlt(x^4)\dlt(x^5), 
 \label{cL:spinor}
\ee
where $\chi$ is a 4D left-handed Weyl spinor, 
and the 2-component spinor~$\psi$ is the 4D right-handed component 
of the 4-component spinor~$\hat{\Psi}$, which is defined as 
\be
 \Psi_+ \equiv \begin{pmatrix} \hat{\Psi} \\ 0 \end{pmatrix}. 
\ee
The 6D gamma matrices~$\Gm^M$ are defined as
\be
 \Gm^\mu = \begin{pmatrix} & \gm^\mu \\ \gm^\mu & \end{pmatrix}, \;\;\;\;\;
 \Gm^4 = \begin{pmatrix} & i\gm_5 \\ i\gm_5 & \end{pmatrix}, \;\;\;\;\;
 \Gm^5 = \begin{pmatrix} & \id_4 \\ -\id_4 & \end{pmatrix}. 
\ee
The brane mass parameter~$c$ is dimensionless, and assumed to be real. 

In the 2-component notation, (\ref{cL:spinor}) is rewritten as
\bea
 \cL \eql -i\lmd\sgm^\mu\der_\mu\bar{\lmd}
 -i\bar{\psi}\bar{\sgm}^\mu\der_\mu\psi
 +\frac{1}{\pi R}\lmd\der_{\bar{z}}\psi
 -\frac{1}{\pi R}\bar{\psi}\der_z\bar{\lmd} \nonumber\\
 &&+\frac{1}{2\pi^2R^2}\brc{-i\chi\sgm^\mu\der_\mu\bar{\chi}
 +c\brkt{\psi\chi+\bar{\psi}\bar{\chi}}}\dlt^{(2)}(z), 
\eea
where $\lmd$ is the 4D left-handed component of $\hat{\Psi}$. 
Thus, the equations of motion are 
\bea
 &&-i\sgm^\mu\der_\mu\bar{\lmd}+\frac{1}{\pi R}\der_{\bar{z}}\psi = 0, 
 \nonumber\\
 &&-i\bar{\sgm}^\mu\der_\mu\psi-\frac{1}{\pi R}\der_z\bar{\lmd}
 +\frac{c}{2\pi^2 R^2}\chi\dlt^{(2)}(z) = 0, 
 \nonumber\\
 &&-i\sgm^\mu\der_\mu\bar{\chi}+c\psi|_{z=0} = 0. 
\eea
From these, we obtain 
\be
 \brc{\Box_4+\frac{1}{\pi^2R^2}\der_z\der_{\bar{z}}
 -\frac{c^2}{2\pi^2R^2}\dlt^{(2)}(z)}\psi = 0. 
 \label{EOM:psi}
\ee
This has the same form as the equation of motion for $\phi$ 
derived from (\ref{cL2}). 
Hence the spectrum in this system is the same as that of the scalar field 
discussed in the text~\cite{Dudas:2005vn}. 

In the case that $\Psi_+$ does not have any charges, 
the following Majorana mass term is allowed on the brane. 
\be
 \cL = i\bar{\Psi}_+\Gm^M\der_M\Psi_+
 +h\brkt{\psi^2+\bar{\psi}^2}\dlt(x^4)\dlt(x^5), 
\ee
where the brane mass parameter~$h$ has the mass dimension~$-1$. 
In contrast to the above Dirac mass term, the equation of motion in this case 
does not have the form of (\ref{EOM:psi}). 
Thus the spectrum in this case has to be discussed separately.

\subsection{Brane mass for vector field}
Here we consider a theory of a vector field~$A^M$. 
We can introduce the following brane mass term. 
\be
 \cL = -\frac{1}{4}F^{MN}F_{MN}
 -\frac{1}{2}\brkt{\der^\mu S+c A^\mu}\brkt{\der_\mu S+c A_\mu}\dlt(x^4)\dlt(x^5), 
 \label{cL:vector}
\ee
where the real scalar~$S$ is the Stueckelberg's scalar field, 
which is localized on the brane. 
The brane mass parameter~$c$ is real and dimensionless. 
The above Lagrangian is invariant under the gauge transformation:
\bea
 A_M(x,z) \toa A_M(x,z)+\der_M\Lmd(x,z), \nonumber\\
 S(x) \toa S(x)-c\Lmd(x,0), 
\eea
where $\Lmd(x,z)$ is the gauge transformation parameter. 

In order to fix the gauge, we add the following gauge-fixing term. 
\be
 \cL_{\rm gf} = -\frac{1}{2\xi}\brc{\der^\mu A_\mu
 +\frac{\xi}{2\pi^2R^2}\brkt{\der_z A_{\bar{z}}+\der_{\bar{z}}A_z}}^2, 
\ee
where $\xi$ is the gauge parameter. 

Performing the partial integral, the total Lagrangian becomes 
\bea
 \cL+\cL_{\rm gf} \eql -\frac{1}{2}\brc{\der^\mu A^\nu\der_\mu A_\nu
 -\brkt{1-\frac{1}{\xi}}\der^\mu A^\nu\der_\nu A_\mu
 +\frac{1}{\pi^2R^2}\der_z A^\mu\der_{\bar{z}}A_\mu} \nonumber\\
 &&-\frac{1}{2\pi^2R^2}\brc{\der^\mu A_z\der_\mu A_{\bar{z}}
 -\frac{1-\xi}{4\pi^2R^2}\brkt{(\der_z A_{\bar{z}})^2+(\der_{\bar{z}}A_z)^2}
 +\frac{1+\xi}{2\pi^2R^2}\abs{\der_z A_{\bar{z}}}^2} \nonumber\\
 &&-\frac{1}{4\pi^2R^2}\brkt{c^2 A^\mu A_\mu-2cS\der^\mu A_\mu
 +\der^\mu S\der_\mu S}\dlt^{(2)}(z).  \label{cL:vector:2}
\eea
We expand the 6D gauge field into the 4D KK modes as
\bea
 A_\mu(x,z) \eql \sum_a u_a(z)A_\mu^{(a)}(x)+\sum_a w_a(z)\der_\mu A_{\rm S}^{(a)}(x), 
 \nonumber\\
 A_z(x,z) \eql \sqrt{2}\pi R\sum_a v_a(z)\vph^{(a)}(x), 
\eea
where $A_\mu^{(a)}(x)$ satisfies $\der^\mu A_\mu^{(a)}(x)=0$. 
The mode functions~$u_a(z)$ and $w_a(z)$ are real, but $v_a(z)$ is complex. 
They are normalized as~\footnote{
Note that $dx^4dx^5=2\pi^2R^2 d^2z$. 
}
\be
 \int_{T^2}d^2z\;u_a^2(z) = \int_{T^2}d^2z\;w_a^2(z) 
 = \int_{T^2}d^2z\;\abs{v_a(z)}^2 = \frac{1}{2\pi^2R^2}. 
\ee
Here we choose the mode functions as
\bea
 u_0(z) \eql w_0(z) = \frac{1}{2\pi R\sqrt{\Im\tau}}, \;\;\;\;\;
 u_{2\hat{a}}(z) = w_{2\hat{a}}(z) = \frac{\cos\brc{2\Im(\lmd_{\hat{a}} z)}}
 {\pi R\sqrt{2\Im\tau}},  \nonumber\\
 u_{2\hat{a}-1}(z) \eql w_{2\hat{a}-1}(z) 
 = \frac{\sin\brc{2\Im(\lmd_{\hat{a}} z)}}{\pi R\sqrt{2\Im\tau}}, \nonumber\\
 v_{n,l}(z) \eql \frac{e^{\lmd_{n,l}z-\lmd_{n,l}^*\bar{z}}}{2\pi R\sqrt{\Im\tau}}, 
\eea
where $a=(n,l)$ (see Sec.~\ref{setup}), $\hat{a}=1,2,\cdots$, and 
\be
 \lmd_{n,l} \equiv \frac{\pi(n+l\bar{\tau})}{\Im\tau}. 
\ee

Then, in terms of the KK modes, (\ref{cL:vector:2}) becomes  
\bea
 \cL^{\rm (4D)} \defa \int_{T^2}dx^4dx^5\;(\cL+\cL_{\rm gf})
 = 2\pi^2R^2\int\dr^2z\;(\cL+\cL_{\rm gf}) \nonumber\\
 \eql \frac{1}{2}\sum_a A^{(a)\mu}\brkt{\Box_4-\frac{\abs{\lmd_a}^2}{\pi^2R^2}}A_\mu^{(a)}
 +\frac{1}{2}\sum_a \der^\mu A_{\rm S}^{(a)}
 \brkt{\frac{\Box_4}{\xi}-\frac{\abs{\lmd_a}^2}{\pi^2R^2}}\der_\mu A_{\rm S}^{(a)} 
 \nonumber\\
 &&-\frac{c^2}{2}\sum_{a,b}u_a(0)u_b(0)\brc{A^{(a)\mu}A_\mu^{(a)}
 +\der^\mu A_{\rm S}^{(a)}\der_\mu A_{\rm S}^{(a)}} \nonumber\\
 &&+c\sum_a u_a(0)S\Box_4 A_{\rm S}^{(a)}
 -\frac{1}{2}\der^\mu S\der_\mu S \nonumber\\
 &&-\sum_a\brc{\der^\mu\vph^{(a)*}\der_\mu\vph^{(a)}
 +\frac{(1+\xi)\abs{\lmd_a}^2}{2\pi^2R^2}\abs{\vph^{(a)}}^2} \nonumber\\
 &&+\sum_{n,l}\frac{1-\xi}{2\pi^2R^2}\Re\brc{\lmd_{n,l}^{*2}\vph^{(n,l)}\vph^{(-n,-l)}}. 
 \label{cL^4D:vector}
\eea
We have used that
\be
 \int\dr^2z\;v_{n,l}(z)v_{n',l'}(z) = \frac{\dlt_{n,-n'}\dlt_{l,-l'}}{2\pi^2R^2}. 
\ee
Furthermore, we decompose $\vph^{(n,l)}(x)$ as
\be
 \vph^{(n,l)} = \frac{e^{i\tht_{n,l}}}{\sqrt{2}}
 \brkt{\vph_{\rm R}^{(n,l)}+i\vph_{\rm I}^{(n,l)}}, 
\ee
where $\tht_{n,l}\equiv\arg(\lmd_{n,l})$. 
Then, the last two lines in (\ref{cL^4D:vector}) is rewritten as
\bea
 \cL^{\rm (4D)}_\vph \eql -\frac{1}{2}\sum_a
 \brc{\der^\mu\vph_{\rm R}^{(a)}\der_\mu\vph_{\rm R}^{(a)}
 +\der^\mu\vph_{\rm I}^{(a)}\der_\mu\vph_{\rm I}^{(a)}
 +\frac{(1+\xi)\abs{\lmd_a}^2}{2\pi^2R^2}
 \brkt{\vph_{\rm R}^{(a)2}+\vph_{\rm I}^{(a)2}}} \nonumber\\
 &&-\frac{1-\xi}{4\pi^2R^2}\sum_{n,l}\abs{\lmd_{n,l}}^2
 \brkt{\vph_{\rm R}^{(n,l)}\vph_{\rm R}^{(-n,-l)}
 -\vph_{\rm I}^{(n,l)}\vph_{\rm I}^{(-n,-l)}}. 
\eea

Thus, the equations of motion are 
\bea
 &&\sum_b\brkt{\Box_4\dlt^{ab}-M_{Aab}^2}A_\mu^{(b)} = 0, \nonumber\\
 &&\Box_4\brc{\sum_b\brkt{\Box_4\dlt^{ab}-\xi M_{Aab}^2}A_{\rm S}^{(b)}
 -\xi c u_a(0)S} = 0, \nonumber\\
 &&\Box_4 S+c\sum_a u_a(0)\Box_4 A_{\rm S}^{(a)} = 0, \nonumber\\
 &&\brc{\Box_4-\frac{(1+\xi)\abs{\lmd_{n,l}}^2}{2\pi^2R^2}}\vph_{\rm R}^{(n,l)}
 -\frac{(1-\xi)\abs{\lmd_{n,l}}^2}{2\pi^2R^2}\vph_{\rm R}^{(-n,-l)} = 0, 
 \nonumber\\
 &&\brc{\Box_4-\frac{(1+\xi)\abs{\lmd_{n,l}}^2}{2\pi^2R^2}}\vph_{\rm I}^{(n,l)}
 +\frac{(1-\xi)\abs{\lmd_{n,l}}^2}{2\pi^2R^2}\vph_{\rm I}^{(-n,-l)} = 0, 
 \label{EOM:vector}
\eea
where
\bea
 M_{Aab}^2 \defa \frac{\abs{\lmd_a}^2}{\pi^2R^2}\dlt_{ab}
 +c^2u_a(0)u_b(0) \nonumber\\
 \eql \begin{cases} \frac{1}{\pi^2R^2}\brkt{\abs{\lmd_a}^2\dlt_{ab}
 +\frac{c^2}{4\Im\tau}} & (a=b=0) \\
 \frac{1}{\pi^2R^2}\brkt{\abs{\lmd_a}^2\dlt_{ab}
 +\frac{c^2}{2\Im\tau}} & (\mbox{$a$ and $b$ : even but $(a,b)=(0,0)$}) \\
 \frac{\abs{\lmd_a}^2}{\pi^2R^2}\dlt_{ab} & (\mbox{$a$ or $b$ : odd}) \end{cases}. 
\eea
From the second and the third equations of (\ref{EOM:vector}), 
we can eliminate $S$ and obtain 
\be
 \Box_4\brkt{\Box_4-\frac{\xi\abs{\lmd_a}^2}{\pi^2R^2}}A_{\rm S}^{(a)} = 0. 
\ee
and from the last two equations of (\ref{EOM:vector}), we obtain 
\bea
 \brkt{\Box_4-\frac{\abs{\lmd_a}^2}{\pi^2R^2}}\vph_{\rm R+}^{(a)} \eql 0, \;\;\;\;\;
 \brkt{\Box_4-\frac{\xi\abs{\lmd_a}^2}{\pi^2R^2}}\vph_{\rm R-}^{(a)} = 0, 
 \nonumber\\
 \brkt{\Box_4-\frac{\xi\abs{\lmd_a}^2}{\pi^2R^2}}\vph_{\rm I+}^{(a)} \eql 0, \;\;\;\;\;
 \brkt{\Box_4-\frac{\abs{\lmd_a}^2}{\pi^2R^2}}\vph_{\rm I-}^{(a)} = 0, 
\eea
where
\bea
 \vph_{\rm R\pm}^{(n,l)} \defa \frac{1}{\sqrt{2}}
 \brkt{\vph_{\rm R}^{(n,l)}\pm\vph_{\rm R}^{(-n,-l)}}, \nonumber\\
 \vph_{\rm I\pm}^{(n,l)} \defa \frac{1}{\sqrt{2}}
 \brkt{\vph_{\rm I}^{(n,l)}\pm\vph_{\rm I}^{(-n,-l)}}. 
\eea

Therefore, we can see that the spectrum for $A_\mu^{(2\hat{a})}(x)$ ($\hat{a}=0,1,\cdots$) 
has similar properties to that of the scalar field discussed in the text. 
In contrast, the spectra for $A_\mu^{(2\hat{a}-1)}(x)$ ($\hat{a}=1,2,\cdots$), 
$\vph_{\rm R+}^{(a)}(x)$ and $\vph_{\rm I-}^{(a)}(x)$ do not receive an effect  
of the brane mass, and are given by 
\be
 m_a = \frac{\abs{\lmd_a}}{\pi R} = \frac{\pi\abs{n+l\tau}}{\pi R\Im\tau}. 
\ee
This is the result from the fact that the mode functions of $A_\mu^{(2\hat{a}-1)}(x)$ 
vanish at the brane, and $A_z(x,z)$ does not have a brane mass in the first place. 
The remaining modes are unphysical, and their spectra depend on the gauge parameter~$\xi$, \ie, 
\be
 m_a = \frac{\sqrt{\xi}\pi\abs{n+l\tau}}{\pi R\Im\tau}. 
\ee
Especially, $\vph_{\rm R-}^{(a)}(x)$ and $\vph_{\rm I+}^{(a)}(x)$ are the would-be NG modes,  
which are absorbed into the longitudinal modes of the massive KK modes for $A_\mu$.


\end{document}